\documentclass{IEEEtran}
\usepackage{cite}
\usepackage{graphicx}
\usepackage{epstopdf}
\usepackage{subfigure}
\usepackage{float}
\usepackage{amssymb,amsfonts,amsmath}
\usepackage{amsfonts,amssymb,bm,color}
\usepackage{amsthm}
\usepackage{amsmath}
\usepackage{algorithm}
\usepackage{algorithmic}
\usepackage{enumerate}
\usepackage{color}
\usepackage{cases}
\usepackage{tabularx} 
\usepackage{booktabs}
\usepackage{diagbox}
\usepackage{array}
\newcommand{\PreserveBackslash}[1]{\let\temp=\\#1\let\\=\temp}
\newcolumntype{C}[1]{>{\PreserveBackslash\centering}p{#1}}
\newcolumntype{R}[1]{>{\PreserveBackslash\raggedleft}p{#1}}
\newcolumntype{L}[1]{>{\PreserveBackslash\raggedright}p{#1}}

\ifCLASSOPTIONcompsoc
\usepackage[caption=false,font=normalsize,labelfon
t=sf,textfont=sf]{subfig}
\else
\usepackage[caption=false,font=footnotesize]{subfi
	g}
\fi

\definecolor{orange}{RGB}{255,107,0}

\makeatletter
\def\endthebibliography{%
	\def\@noitemerr{\@latex@warning{Empty `thebibliography' environment}}%
	\endlist
}
\makeatother
\begin{document}
	\bibliographystyle{IEEEtran}
\title{
{Cooperative Rate-Splitting for MISO Broadcast Channel with User Relaying, and Performance Benefits over Cooperative NOMA}
}
\author{Jian Zhang, Bruno Clerckx, {\IEEEmembership{Senior Member, IEEE}}, Jianhua Ge, and Yijie Mao
\thanks{
\vspace{-0.0cm}
J. Zhang and J. Ge are with the ISN State Key Laboratory, Xidian University, Xi'an, China (e-mail: j.zhang18@imperial.ac.uk, jhge@xidian.edu.cn).
B. Clerckx is with Imperial College London, London SW7 2AZ, UK
(email: b.clerckx@imperial.ac.uk). Y. Mao is with The University of Hong
Kong, Hong Kong, China (email: maoyijie@eee.hku.hk). This work has been partially supported by the EPSRC of the UK under grant EP/N015312/1.
}
\vspace{-0.6cm}
}
\maketitle

\vspace{-0.6cm}
\begin{abstract}
Due to its promising performance in a wide range of practical scenarios, Rate-Splitting (RS) has recently received significant attention in academia for the downlink of communication systems.  In this letter, we propose and analyse a Cooperative Rate-Splitting (CRS) strategy based on the three-node relay channel where the transmitter is equipped with multiple antennas. By splitting user messages and linearly precoding common and private streams at the transmitter, and opportunistically asking the relaying user to forward its decoded common message, CRS can efficiently cope with a wide range of propagation conditions (disparity of user channel strengths and   directions) and compensate for the performance degradation due to deep fading. 
The precoder design and the resource allocation  are optimized by solving the Weighted Sum Rate (WSR)  maximization problem.
Numerical results demonstrate that our proposed CRS scheme can achieve an explicit rate region improvement compared to its non-cooperative counterpart and other cooperative strategies (such as cooperative NOMA). 
\end{abstract}

\vspace{-0.1cm}
\section{Introduction}

\IEEEPARstart{T}{he} classical three-node relay channel (or cooperative network) was first introduced by Van Der Meulen\cite{Meulen1971}. 
Due to its superior ability to improve the reliability, the relay channel has been extensively studied from various aspects, including cooperative diversity, capacity, and resource allocation \cite{Liang2005,Liang2007,Gunduz2007,Mesbah2008,Host-Madsen2005}. 
Of particular interest is the cooperative Relay Broadcast Channel (RBC)\cite{Liang2007a,Liang2007b}. In the partially cooperative two-user RBC, user 1 acts as a relay node and forwards cooperative information to user 2 through a relay-aided link. Such strategy is helpful in the deployments where user 1 experiences a better channel than user 2 and may decode information intended for user 2 in addition to its own information. In the special case where the partially cooperative RBC is degraded, the rate region achieved by user cooperation is shown to be the capacity region\cite{Liang2007a}. This idea has recently been revived in the context of cooperative Non-Orthogonal Multiple Access (NOMA). As NOMA\footnote{There exist different forms of NOMA strategies. In this letter, we only refer to power-domain NOMA and simply use the terminology ``NOMA''.} similarly leverages the user channel disparity, cooperative NOMA is capable to  improve the outage performance, as well as the spectral and energy efficiencies over orthogonal multiple access in the degraded RBC\cite{Ding2015,Wan2018,Kim2015,Liu2017}.           

However, although NOMA based on Superposition Coding (SC) and Successive Interference Cancellation (SIC) performs well in degraded Single-Input Single-Output 
Broadcast Channel (SISO BC), it faces several bottlenecks in a multi-antenna scenario, such as Degrees-of-Freedom (DoF) loss, and performance degradation in general user deployments \cite{Mao2018a,Bruno2019}. In contrast to NOMA that relies on some users to fully decode the messages of other users (i.e. fully decode interference), Rate-Splitting (RS) also exploits SIC but exploits a more flexible framework of non-orthogonal transmission, which enables to partially decode interference and partially treat remaining interference as noise, and hence provide significant benefits in terms of spectral efficiency\cite{Mao2018a,Hao2015,Joudeh2016,Dai2016,Lu2018,Joudeh2017,Mao2018c}, energy efficiency\cite{Mao2018b}, robustness\cite{Joudeh2016a}, and CSI feedback overhead reduction\cite{Hao2015,Dai2017}. In addition, on the standpoint of encoding structure, the combination of a common stream decoded by all users and private streams dedicated to each user also provides a fundamental compatibility with other advanced techniques\cite{Clerckx2016}. 
Such promising compatibility further motivates us to investigate the application of RS in cooperative relaying.

In this letter, a cooperative strategy is proposed to naturally link the multi-antenna RS technique, developed for multi-antenna BC, and a three-node relay broadcast channel with one transmitter and two receivers. The proposed scheme, denoted as Cooperative RS (CRS), relies on user cooperation such that the rate of the common stream in RS can be enhanced. 
The transmitter splits user messages into common and private parts, encodes them into common and private streams and linearly precodes them before transmission. One user then decodes the common stream and forwards it to the other user, therefore helping that user to better decode the common stream. After performing SIC and removing the common stream, both users decode their respective private stream, and then reconstruct their original intended message from the private and common streams.
With the proposed flexible framework,
CRS leverages the benefits not only of RS for multi-antenna BC \cite{Mao2018a,Bruno2019,Joudeh2016,Hao2015}, but also of user cooperation \cite{Liang2005,Liang2007,Gunduz2007,Liang2007a,Liang2007b}. 
For some channel conditions, the CRS framework boils down to Non-cooperative RS (NRS) whenever user cooperation is not needed, and encompasses cooperative NOMA as a special case. 
The precoder design and resource allocation problem with the objective of maximizing the Weighted Sum Rate (WSR)  are further investigated.
A modified Weighted Minimum Mean Square Error (WMMSE) approach is proposed to solve the challenging non-convex problem by using Alternating Optimization (AO). Simulation results demonstrate that our proposed CRS is more flexible than NRS, cooperative NOMA,  fixed
equal-time cooperative RS, and can achieve an explicit rate region improvement compared to these baseline schemes in a wide range of propagation conditions. 


\vspace{-0.2cm}
\begin{figure}[!t]
	\centering
	\includegraphics[width=1\linewidth]{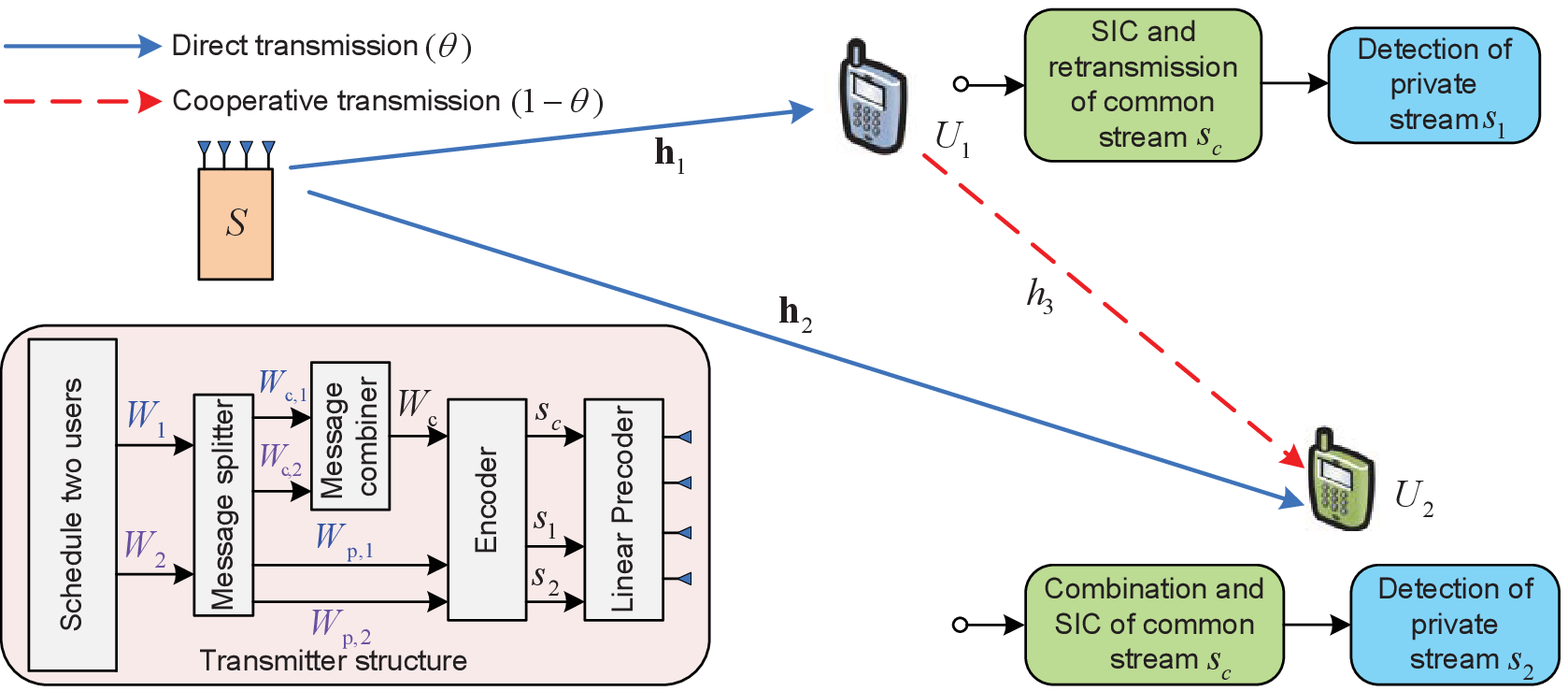}
	\caption{Cooperative Rate-Splitting scheme.}
	\label{fig1}
\end{figure}
\vspace{-0.2cm}
\section{System Model}
Consider a three-node downlink cooperative network as shown in Fig. 1, where a base station $S$ equipped with $N_T$ (${N_T} \ge 2$) transmit antennas intends to communicate with two single-antenna users $U_1$ and $U_2$.  $U_1$ is regarded as user relaying and the Non-regenerative Decode-and-Forward (NDF) protocol \cite{Mesbah2008} is employed to help the transmission of $U_2$. Three wireless channels at links $S \to U_1$,  $S \to U_2$, $U_1 \to U_2$ are denoted as ${\bf{h}}_1$,  ${\bf{h}}_2$, and  ${h}_3$, respectively. Following the RS principle for multi-antenna BC\cite{Clerckx2016}, the message $W_k$ intended to user $k$, $\forall k \in \{ {1,{\rm{ }}2} \}$, is split into a common part ${W_{c,k}}$ and a private part ${W_{p,k}}$.
The common parts ${W_{c,1}}$  and ${W_{c,2}}$  are encoded into the stream $s_c$  using a common codebook, and hence is decoded by both users. ${W_{p,1}}$  and ${W_{p,2}}$ are encoded into private streams $s_1$ and $s_2$, respectively. The three streams ${\bf{s}} = [ {{s_c},{s_1},{s_2}} ]$  are linearly precoded using   ${\bf{P}} = [ {{\bf{p}}_c,{{\bf{p}}_1},{\bf{p}}_2} ]$, where $[{\bf{p}}_c, {{\bf{p}}_1},{\bf{p}}_2] \in {\mathbb{C}^{{N_T} \times 3}}$.
The resulting transmit signal is 
\vspace{-0.1cm}
\begin{equation}\label{1}
{\bf{x}} = {\bf{Ps}} = {\bf{p}}_c{s_c} + {\bf{p}}_1{s_1} + {\bf{p}}_2{s_2}\text{.}
\end{equation}
Assuming that ${\mathbb{E}}\{ {{\bf{s}}{{\bf{s}}^H}} \} \!=\!\bf{I}$, the  transmit power constraint at $S$ is given as  ${\text{tr}}( {{\bf{P}}{{\bf{P}}^H}} ) \le {P_t}$. 

In this work, we assume that the relaying user operates in the Half-Duplex (HD) mode where the communication process is completed in two consecutive slots. Different from the conventional HD equal-time allocation strategies, we further introduce a parameter $\theta $  ($0 < \theta  \le 1$) to represent the channel resource allocation between the two time-domain orthogonal channels.  More specifically, for a given total time, the fraction of time $\theta$ is allocated to the direct transmission ($S\rightarrow U_1$, $S\rightarrow U_2$), and the remaining fraction $(1 - \theta )$ is allocated to the cooperative transmission ($U_1\rightarrow U_2$).

In the direct transmission (or first) slot, the base station sends the superposed signal $\bf{x}$ to $U_1$ and $U_2$ simultaneously. 
Therefore, the received signal at user $k$ in the first slot is written as 
\begin{equation}\label{2}
y_k^{(1)} = {\bf{h}}_k^H{\bf{x}} + {n_k}\text{,}
\end{equation}
where ${n_k}\!\! \sim \!\!{\cal{CN}}(0,\sigma _{k}^2)$ is the Additive White Gaussian Noise (AWGN). Without loss of generality, we assume that Channel State Information\footnote{Although the superiority of RS under imperfect CSIT has been extensively discussed in \cite{Joudeh2016,Hao2015,Joudeh2016a}, RS has also been shown to be useful in perfect CSIT settings \cite{Mao2018a,Joudeh2017,Bruno2019}, and our objective of this work is to obtain some insightful results about the application of RS in cooperative relaying. Therefore, we only focus on the perfect CSIT case here, and the analysis of imperfect CSIT is beyond the scope of this letter.} (CSI) of all links are known perfectly at $S$, and the noise variances across users are equal to one ($\sigma_k^2=1$). 

At the each user, the common stream $s_c$  is  decoded 
while treating all the other streams as noise. Hence, the instantaneous signal to interference plus noise ratio (SINR) of decoding  $s_c$ at user $k$ is 
\vspace{-0.2cm}
\begin{equation}\label{3}
\gamma _{c,k}^{(1)} = \frac{{{{\left| {{\bf{h}}_k^H{{\bf{p}}_c}} \right|}^2}}}{{{{\left| {{\bf{h}}_k^H{{\bf{p}}_1}} \right|}^2} + {{\left| {{\bf{h}}_k^H{{\bf{p}}_2}} \right|}^2} + 1}}\text{.}
\vspace{-0.1cm}
\end{equation}

During the cooperative (or second) slot, $U_1$ re-encodes the decoded common stream $s_c$ using a different codebook generated independently from that of $S$, and then forwards it to $U_2$ in a power level $P_R$, while $S$ remains silent. As a result, the received signal from the SISO relay-aided path at $U_2$ is given by  $y_2^{(2)} = {h_3}\sqrt {{P_R}} {s_c} + {n_3}$, and the corresponding SINR is  $\gamma _{c,2}^{(2)} = {| {{h_3}\sqrt {{P_R}} } |^2}$.


Following a strategy similar to \cite{Liang2005,Liang2007,Gunduz2007,Mesbah2008,Host-Madsen2005}, $U_2$ then combines the decoding results of the two slots for $s_c$. To ensure that $s_c$ can be decoded successfully by both $U_1$ and $U_2$, the achievable rate of the common stream is given by
\vspace{-0.1cm}
\begin{equation}\label{5}
\begin{array}{l}
R_c = \min \big[\theta \log_2(1 + \gamma _{c,1}^{(1)}), \theta \log_2(1 + \gamma _{c,2}^{(1)}) \\
\quad\quad\ + (1 - \theta )\log_2(1 + \gamma _{c,2}^{(2)}) \big]\text{.}
\end{array}
\vspace{-0.1cm}
\end{equation}
After performing the SIC and removing the common stream, the SINR of decoding private stream $s_k$  at user $k$ is 
\vspace{-0.1cm}
\begin{equation}\label{4}
	\gamma _k^{(1)} = \frac{{{{\left| {{\bf{h}}_k^H{{\bf{p}}_k}} \right|}^2}}}{{{{\left| {{\bf{h}}_k^H{{\bf{p}}_j}} \right|}^2} + 1}}\text{,}
	\vspace{-0.1cm}
\end{equation}
where  $k,j \in 1,2$ and  $k \ne j$. Consequently, the achievable rate of private stream $s_k$ is given as ${R_{p,k}} = \theta{\text{log}}{_2}(1 + \gamma _k^{(1)})$. 

In this letter, we focus on the WSR maximization problem in this HD cooperative network. 
For a given pair of user weights ${\bf{u}} = [ {{u_1},{u_2}} ]$ and a given time allocation factor  $\theta$, the WSR optimization problem is formulated as:
\vspace{-0.1cm}
\begin{subnumcases}{R_{\mathrm{WSR}}({\bf{u}},{P_t},{P_R},\theta)\!\!:\!\!} 
	\!\!\mathop {\max }\limits_{\bf{P, c}} {u_1} {R_{1,{\rm{tot}}}}\!  +\!  {u_2}  {R_{2,{\rm{tot}}}} \\
	\vspace{-0.2cm}
	{\text{s.t.}}\ {\rm{ }}{{\rm{c}}_{\rm{1}}} + {R_{p,1}} \ge R_1^{\text{tar}} \\
	\!\ \ \quad {{\rm{c}}_{\rm{2}}} +  {R_{p,2}} \ge R_2^{{\text{tar}}}\\
	\!\ \ \quad {{\rm{c}}_{\rm{1}}} + {{\rm{c}}_{\rm{2}}} \le {R_c}\\
	\!\ \ \quad {{\rm{c}}_{\rm{1}}} \ge 0,{\rm{ }}{{\rm{c}}_{\rm{2}}} \ge 0\\
	\!\ \ \quad  {\text{tr}}\{ {\bf{P}}{{\bf{P}}^{\rm{H}}}\}  \le {P_t}{\rm{ }}
\end{subnumcases}
where ${{\rm{c}}_{\rm{k}}}$ denotes the rate of the common part of the $k$-th user's message (i.e., $W_{c,k}$), and $\mathbf{c}=[{{\rm{c}}_{\rm{1}}}, {{\rm{c}}_{\rm{2}}}]$.  ${R_{k,{\rm{tot}}}}={R_{p,k}} + {{\rm{c}}_{\rm{k}}}$ denotes the total rate of user $k$. (6b) and (6c) are the Quality-of-Service (QoS) constraints. $R_1^{{\text{tar}}}$  and $R_2^{{\text{tar}}}$ denote the individual rate constraints of $U_1$ and $U_2$, respectively. (6d) is the common rate constraint.  
Constraint (6e) implies the corresponding rate is non-negative. (6f) is the transmit power constraint at $S$.  

The CRS scheme described above offers a flexible formulation, and encompasses several conventional  schemes as special cases. Specifically, by adjusting the parameter $\theta$ to 1, CRS switches to conventional NRS, thus retaining all the superiority  of NRS over Non-cooperative NOMA (N-NOMA) and Multi-User Linear Precoding\footnote{As for the comprehensive comparisons between NRS, N-NOMA and MU-LP in various scenarios, and how NRS generalizes and subsumes several relevant schemes as special cases, readers are referred to \cite{Mao2018a,Bruno2019}.} (MU-LP) naturally. By adjusting $\theta$ to $1/2$, CRS reduces to the common fixed Equal-time Cooperative RS (ERS), which is easy to implement in practice. Additionally, by encoding the entire $W_2$ into $s_c$  while allocating no power to  ${\bf{p}}_2$, i.e.,  ${\bf{x}}= {{\bf{p}}_c}{s_c} + {{\bf{p}}_1}{s_1}$ with flexible $\theta$,
CRS boils down to a multi-antenna version of cooperative NOMA (C-NOMA) with dynamic time allocation. Moreover, by switching off ${\bf{p}}_k$ and encoding the entire $W_2$ into $s_c$, i.e., ${\bf{x}}= {{\bf{p}}_c}{s_c}$  with flexible $\theta$, Single-user Opportunistic Decode-and-Forward Transmission (ODF) in \cite{Gunduz2007} is directly obtained as an extreme case where $U_1$ is served as a dedicated relay. The mapping of messages to streams, introduced in\cite{Bruno2019} for non-cooperative settings, is further extended to the cooperative setting in Table 1 for several subset schemes.
  
\begin{table}[!t]
	\caption{Mapping of messages to streams.}
	\setlength{\tabcolsep}{0.0mm}{
		\begin{tabular}{|L{1.6cm}|C{1.3cm}|C{1.3cm}|C{2.3cm}|C{1.7cm}|}
			\hline
			$ $&$s_1$&$s_2$&$s_c$&\scriptsize$\theta$\\
			\cline{1-4}
			\hline
			\ N-NOMA&\textcolor[rgb]{0,0.5,0.7}{{\scriptsize{$W_1$}}}&-&\textcolor{red}{\scriptsize$W_2$}&\scriptsize$\theta=1$\\
			\hline
			\ MU-LP&\textcolor[rgb]{0,0.5,0.7}{\scriptsize$W_1$}&\textcolor[rgb]{0,0.5,0.7}{\scriptsize$W_2$}&-&\scriptsize$\theta=1$\\
			\hline
			\ NRS&\textcolor[rgb]{0,0.5,0.7}{\scriptsize$W_{p,1}$}&\textcolor[rgb]{0,0.5,0.7}{\scriptsize$W_{p,2}$}&\textcolor{red}{\scriptsize$W_{c,1}$, $W_{c,2}$}&\scriptsize$\theta=1$\\
			\hline
			\ ERS&\textcolor[rgb]{0,0.5,0.7}{\scriptsize$W_{p,1}$}&\textcolor[rgb]{0,0.5,0.7}{\scriptsize$W_{p,2}$}&\textcolor{red}{\scriptsize$W_{c,1}$, $W_{c,2}$}&\scriptsize$\theta=1/2$\\
			\hline
			\ ODF&-&-&\textcolor{red}{\scriptsize$W_{2}$}&\scriptsize$0<\theta<1$\\
			\hline
			\ C-NOMA&\textcolor[rgb]{0,0.5,0.7}{\scriptsize$W_1$}&-&\textcolor{red}{\scriptsize$W_2$}&\scriptsize$0<\theta<1$\\
			\hline
			\ CRS&\textcolor[rgb]{0,0.5,0.7}{\scriptsize$W_{p,1}$}&\textcolor[rgb]{0,0.5,0.7}{\scriptsize$W_{p,2}$}&\textcolor{red}{\scriptsize$W_{c,1}$,  $W_{c,2}$}&\scriptsize$0<\theta<1$\\
			\hline
			\multicolumn{1}{c}{ }&\multicolumn{2}{p{2.7cm}<{\centering}}{\textcolor[rgb]{0,0.5,0.7}{decoded only by its intended user}}&\multicolumn{1}{p{2.7cm}<{\centering}}{\textcolor{red}{decoded by both users}}\\ 	
		\end{tabular}}	
		
		\vspace{-6mm}
	\end{table}
	
\section{Optimization Framework}
Since the problem in (6) is non-convex and very challenging to solve, in this section, we propose an AO algorithm based on modified WMMSE approach to jointly design the beamforming vectors and time-allocation parameter. 

Although the Mean Square Error (MSE) expressions and Minimum MSE (MMSE) receivers are known in the literature \cite{Christensen2008}, below we start by rewriting them for coherent presentation. Let ${\widehat s_{c,k}} = {g_{c,k}}y_k^{(1)}$  denote the user $k$'s estimate of ${s_c}$ in the direct transmission, where ${g_{c,k}}$  is a scalar equalizer. After SIC, the estimate of $s_k$ is obtained as  ${\widehat s_k} = {g_k}(y_k^{(1)} - {\bf{h}}_k^H{{\bf{p}}_c}{\widehat s_{c,k}})$, where  $g_k$ is the equalizer. The MSE of the common stream at user $k$ is defined as  ${\varepsilon _{c,k}} \triangleq \mathbb{E}\{ {{{| {{{\widehat s}_{c,k}} - {s_c}} |}^2}} \}$, and the MSE of the private stream is defined as 
$\varepsilon_k=\mathbb{E}\{|\widehat{s}_k-s_k|^2\}$. Letting ${T_{c,k}} = {\left| {{\bf{p}}_c^H{{\bf{h}}_k}} \right|^2} + {T_k}$ and  ${T_k} = \sum\nolimits_{i = 1}^2 {{{\left| {{\bf{p}}_i^H{{\bf{h}}_k}} \right|}^2}}  + 1$, the MSEs can be expressed as:
\vspace{-0.2cm}
\begin{equation}\label{7}
\vspace{-0.1cm}
\begin{array}{l} 
{\varepsilon _{c,k}} = {\left| {{g_{c,k}}} \right|^2}{T_{c,k}} - 2\Re \left\{ {{g_{c,k}}{\bf{h}}_k^H{{\bf{p}}_c}} \right\} + 1\text{,}\\
{\varepsilon _k} = {\left| {{g_k}} \right|^2}{T_k} - 2\Re \left\{ {{g_k}{\bf{h}}_k^H{{\bf{p}}_k}} \right\} + 1\text{.}
\end{array}
\vspace{-0.1cm}
\end{equation}
By solving $\frac{{\partial {\varepsilon _{c,k}}}}{{\partial {g_{c,k}}}} = 0$  and  $\frac{{\partial {\varepsilon _{k}}}}{{\partial {g_{k}}}} = 0$, the well-known MMSE equalizers are obtained as $g_{c,k}^{{\text{MMSE}}} = {\bf{p}}_c^H{{\bf{h}}_k}T_{c,k}^{ - 1},{\text{  and }}g_k^{{\text{MMSE}}} = {\bf{p}}_k^H{{\bf{h}}_k}T_k^{ - 1}$.
Substituting the equalizers into (7), the MMSEs are given as $\varepsilon _{c,k}^{{\text{MMSE}}}\! = \!\mathop {\min}\limits_{{g_{c,k}}}{\varepsilon _{c,k}}\! =\! T_{c,k}^{ - 1}{I_{c,k}},{\text{ }}\varepsilon _k^{{\text{MMSE}}}\! =\! \mathop {\min}\limits_{{g_k}}{\varepsilon _k}\! =\! T_k^{ - 1}{I_k}\text{,}$
where  ${I_{c,k}} = {T_k}$ and  ${I_k} = {T_k} - {| {{\bf{p}}_k^H{{\bf{h}}_k}} |^2}$. 

The MMSEs and the SINRs in (3) and (4) are related such that $\gamma _{c,k}^{(1)} = 1/\varepsilon _{c,k}^{{\text{MMSE}}} - 1$ and $\gamma _k^{(1)} = 1/\varepsilon _k^{{\text{MMSE}}} - 1$. Consequently, the achievable private rate of user $k$ is given by ${R_{p,k}} =  - {\log _2}(\varepsilon _k^{{\text{MMSE}}})$, and the corresponding common rate is given by  $R_{c,k}^{(1)} =  - {\log _2}(\varepsilon _{c,k}^{{\text{MMSE}}})$. 
The augmented WMSEs \cite{Christensen2008} are expressed as ${\xi _{c,k}}\! = \!{w_{c,k}}{\varepsilon _{c,k}}\!\! -\!\! {\log _2}({w_{c,k}}),{\text{ and }}{\xi _k}\! =\! {w_k}{\varepsilon _k} \!-{\log _2}({w_k})\text{,}$ where ${w_{c,k}}$  and $w_k$  are the weights associated with the user $k$'s MSEs.

From $\frac{{\partial {\xi _{c,k}}}}{{\partial {g_{c,k}}}} = 0$ and $\frac{{\partial {\xi _k}}}{{\partial {g_k}}} = 0$, the optimum equalizers are given as $g_{c,k}^ *  = g_{c,k}^{{\text{MMSE}}}$ and  $g_k^ *  = g_k^{{\text{MMSE}}}$. 
Then, the augmented WMSEs are given by: 
\vspace{-0.1cm}
\begin{equation}\label{11}
\begin{array}{l} 
{\xi _{c,k}}(g_{c,k}^{{\text{MMSE}}}) = {w_{c,k}}\varepsilon _{c,k}^{{\text{MMSE}}} - {\log _2}({w_{c,k}}),\\
{\xi _k}(g_k^{{\text{MMSE}}}) = {w_k}\varepsilon _k^{{\text{MMSE}}} - {\log _2}({w_k})\text{.}
\end{array}
\vspace{-0.1cm}
\end{equation}
Subsequently, from $\frac{{\partial {\xi _{c,k}}(g_{c,k}^{{\text{MMSE}}})}}{{\partial {w_{c,k}}}} = 0$ and  $\frac{{\partial {\xi _k}(g_k^{{\text{MMSE}}})}}{{\partial {w_k}}} = 0$, the optimum MMSE weights can be obtained as  
\vspace{-0.1cm}
\begin{equation}\label{12}
w_{c,k}^ *\!\!  = \!\!w_{c,k}^{{\text{MMSE}}} \!\!\triangleq\!\! {\left( {\varepsilon _{c,k}^{{\text{MMSE}}}} \right)\!\!^{ - 1}}\!\!\!\!,{\text{ and }}w_k^ * \!\! =\!\! w_k^{{\text{MMSE}}} \!\!\triangleq\!\! {\left( {\varepsilon _k^{{\text{MMSE}}}} \right)\!\!^{ - 1}}\text{.}
\vspace{-0.1cm}
\end{equation}
Therefore, the corresponding Rate-WMMSE relationship is established as
\vspace{-0.1cm}
\begin{equation}\label{13}
\xi _{c,k}^{{\text{MMSE}}}\!\! \triangleq\!\!\!\!\! \mathop {\min }\limits_{{w_{c,k}},{g_{c,k}}}\!\!\! {\xi _{c,k}}\!\! =\!\! 1\! -\!R_{c,k}^{(1)}{\text{, }}\xi _k^{{\text{MMSE}}}\!\! \triangleq\!\!\! \mathop {\min }\limits_{{w_k},{g_k}}\!\! {\xi _k}\!\! =\!\! 1\! -\! R_{p,k}\text{.}
\vspace{-0.3cm}
\end{equation}

Next, we demonstrate how to transform the original problem in (6) to the equivalent WMMSE problem.  
With formula (10), the objective function in (6a) can be modified as $\mathop {\max }\limits_{\bf{P, c}} {\rm{ }}{u_1}\theta (1 - {\xi _1}) + {u_2}\theta (1 - {\xi _2}) + {u_1}{{\rm{c}}_1} + {u_2}{{\rm{c}}_2}$. For brevity, $\xi_{c,k}$ and $\xi_k$ are referred to the augmented WMSEs in (8). Denoting ${R_{c,2}^{(2)}} = \log_2(1 + \gamma _{c,2}^{(2)})$  as the achievable rate of relay-aided link, the common rate constraint (6d) can be rewritten as ${{\rm{c}}_1} + {{\rm{c}}_2} \le \min \{ {\theta (1 - {\xi _{c,1}}),{\rm{ }}\theta (1 - {\xi _{c,2}}) + (1 - \theta ){R_{c,2}^{(2)}}} \}$. Therefore,  we can reformulate the modified WMMSE problem for a given ${\bf{u}} = \left[ {{u_1},{u_2}} \right]$ and a given  $\theta$  as:
\vspace{-0.1cm}
\begin{equation}\label{15}
\vspace{-0.1cm}
\begin{array}{l}
{\rm{      }}\mathop {\min }\limits_{\bf{p, \overline c , g,{\rm{ }}w}} {\rm{ }}{u_1}\theta {\xi _1} + {u_2}\theta {\xi _2} + {u_1}{\overline {\rm{c}} _1} + {u_2}{\overline {\rm{c}} _2}\\
\qquad{\text{s.t.}}\ {\rm{ }}{\overline {\rm{c}} _1} + {\overline {\rm{c}} _2} \ge {\xi _c} - \theta \\
\quad\quad\quad\ {\overline {\rm{c}} _1} \le 0,{\rm{ }}{\overline {\rm{c}} _2} \le 0\\
\quad\quad\quad\  - {\overline {\rm{c}} _1} + \theta (1 - {\xi _1}) \ge R_1^{{\text{tar}}}\\
\quad\quad\quad\  - {\overline {\rm{c}} _2} + \theta (1 - {\xi _2}) \ge R_2^{{\text{tar}}}\\
\quad\quad\quad\ {\text{tr}}\{ {\bf{P}}{{\bf{P}}^{\rm{H}}}\}  \le {P_t}
\end{array}
\vspace{-0.2cm}
\end{equation}
where  ${\xi _c} = \max \{ {\theta {\xi _{c,1}},{\rm{ }}\theta {\xi _{c,2}} - (1 - \theta ){R_{c,2}^{(2)}}} \}$, ${\bf{\overline c}}  = [{\overline {\text{c}} _1}, {\overline {\text{c}} _2}] = [ - {{\text{c}}_1}, - {{\text{c}}_2}]$  is the transformation of the common stream rate,  ${\bf{g}} = [g_1, g_2, g_{c,1}, g_{c,2}]$ and ${\bf{w}} = [w_1, w_2, w_{c,1}, w_{c,2}]$  are the equalizers and weights associated with MSEs, respectively.  

By minimizing (11) with respect to ${\bf{g}}$  and  ${\bf{w}}$, the optimal MMSE equalizers and weights can be obtained as 
${{\bf{g}}^{{\text{MMSE}}}}$ and ${{\bf{w}}^{{\text{MMSE}}}}$,
which meet the KKT conditions for a given  ${\bf{P}}$. Furthermore, with (10), if the solution ($
\bf{P^ * },\overline{c}^*,{g^ * },{w^ * }$) satisfies the KKT optimality conditions of (11), the solution ($\bf{P^ * },{c^*}=-\overline{c}^*$) also satisfies the KKT optimality conditions of (6). Therefore, the original problem in (6) and minimum WMMSE problem in (11) are equivalent and it is sufficient to design optimum precoders for only one of the problems.

Although the joint optimization of ($\bf{P,\overline c ,g,w}$) in problem (11) is non-convex, it is convex in each of the blocks. With a given ($\bf{P,\overline c}$), the closed-form of ${{\mathbf{g}}^{{\text{MMSE}}}}$ and ${{\mathbf{w}}^{{\text{MMSE}}}}$   are optimal. With a given ($\bf{g,w}$), problem (11) is a convex Quadratically Constrained Quadratic Program (QCQP) which can be solved using interior-point methods. 

Based on the properties described above, we exploit the AO algorithm to solve the problem, which is shown as follows.
\vspace{-0.1cm}
\noindent\rule{\linewidth}{0.5mm}
\vspace{-0.1cm}
{Alternating Optimization Algorithm}
\vspace{-0.15cm}

\noindent\rule{\linewidth}{0.5mm}
\begin{enumerate}[1:]
	\item {\bf{Input}}:${\bf{u}},{P_t},{P_R},\theta $,	
	 {\bf{Initialize}}: $n\!\! \leftarrow\! 0$, $R_{\mathrm{WSR}}^{[n]}(\theta)\!\! \leftarrow\!\! 0$,  ${{\bf{P}}^{\left[ n \right]}}$ with the MRT-SVD approach of [16, Sec. VI]
	\item {\bf{repeat}}:  $n\! \leftarrow\! n + 1$, ${{\bf{P}}^{[n - 1]}}\! \leftarrow\! {{\bf{P}}}$
	\item \qquad    ${\bf{g}}\leftarrow {{\bf{g}}^{{\text{MMSE}}}}({{\bf{P}}^{[n - 1]}})$, ${\bf{w}} \leftarrow {{\bf{w}}^{{\text{MMSE}}}}({{\bf{P}}^{[n - 1]}})$
	\item  \qquad   update ($\bf{P,\overline c }$) by solving (11) with the updated $\bf{g}$ and 
	
     \qquad  $\bf{w}$, update $R_{\mathrm{WSR}}^{[n]}$  with updated ($\bf{P,\overline c }$)
	\item {\bf{until}} $\left| {R_{\mathrm{WSR}}^{[n]}(\theta ) - R_{\mathrm{WSR}}^{[n - 1]}(\theta )} \right| < \varepsilon$  
	\item {\bf{Output}}:${R_{\mathrm{WSR}}}(\theta )$, $\bf{P}(\theta )$, and $\bf{\overline c}(\theta )$ 
\end{enumerate}

\vspace{-0.3cm} 
\noindent\rule{\linewidth}{0.5mm}
As the output $\mathbf{P}(\theta)$ and $\bf{\overline c}(\theta )$ are derived for a given $\theta$, (6a) becomes a univariate function of  $\theta$. In addition, as $\theta$ is a one-dimensional variable, we can search for the optimal $\theta$ in the range of  $( {0,{\rm{ }}1} ]$, and use the AO algorithm to obtain the precoder set $\bf{P}$ for each specific $\theta$.  The corresponding ($\theta, {{\bf{P}}, {\bf{\overline c}}}$) that jointly maximize the objective (6a) is the final solution. 
\vspace{-0.1cm}
\section{Numerical Results and Discussions}
In this section, 
to comprehensively assess the benefits of CRS, five existing beamforming schemes are compared as baselines, namely, NRS, ERS, N-NOMA, MU-LP, and C-NOMA with the optimal time-resource allocation. In this analysis, the rate region serves as the key performance indicator as it conveys information about the achievable rate by each user under different service requirement priorities. 
Following a similar setup as in \cite{Mao2018a,Mao2018c}, we consider the specific channel realizations to investigate the effect of several channel parameters. When  ${N_T} = 4$, the channels 
are given by  ${{\bf{h}}_1} = [1,1,1,1]^H$,  ${{\bf{h}}_2} = {\lambda _1} \times [1,{e^{j\alpha }},{e^{j2\alpha }},{e^{j3\alpha }}]^H$, and  ${h_3} = {\lambda _2}$, respectively, where ${\lambda _1}$   and ${\lambda _2}$  control the relative channel strength,  $\alpha$ controls the channel angle between $U_1$ and $U_2$. We assume that  ${P_t} = {P_R}$,  $R_k^{\text{tar}} = 0$, and ${{{u}}_1}$ is always equal to 1 while the weight for $U_2$ is varied as ${{\bf{u}}_2} = {10^{[ - 3, - 1:0.05:1,3]}}$ corresponding to a total of 43 different weights. 
\vspace{-0.1cm}
\begin{figure}[!h]
	\vspace{-3mm}
	\centering
	\vspace{-0.05cm}
	\includegraphics[width=1\linewidth]{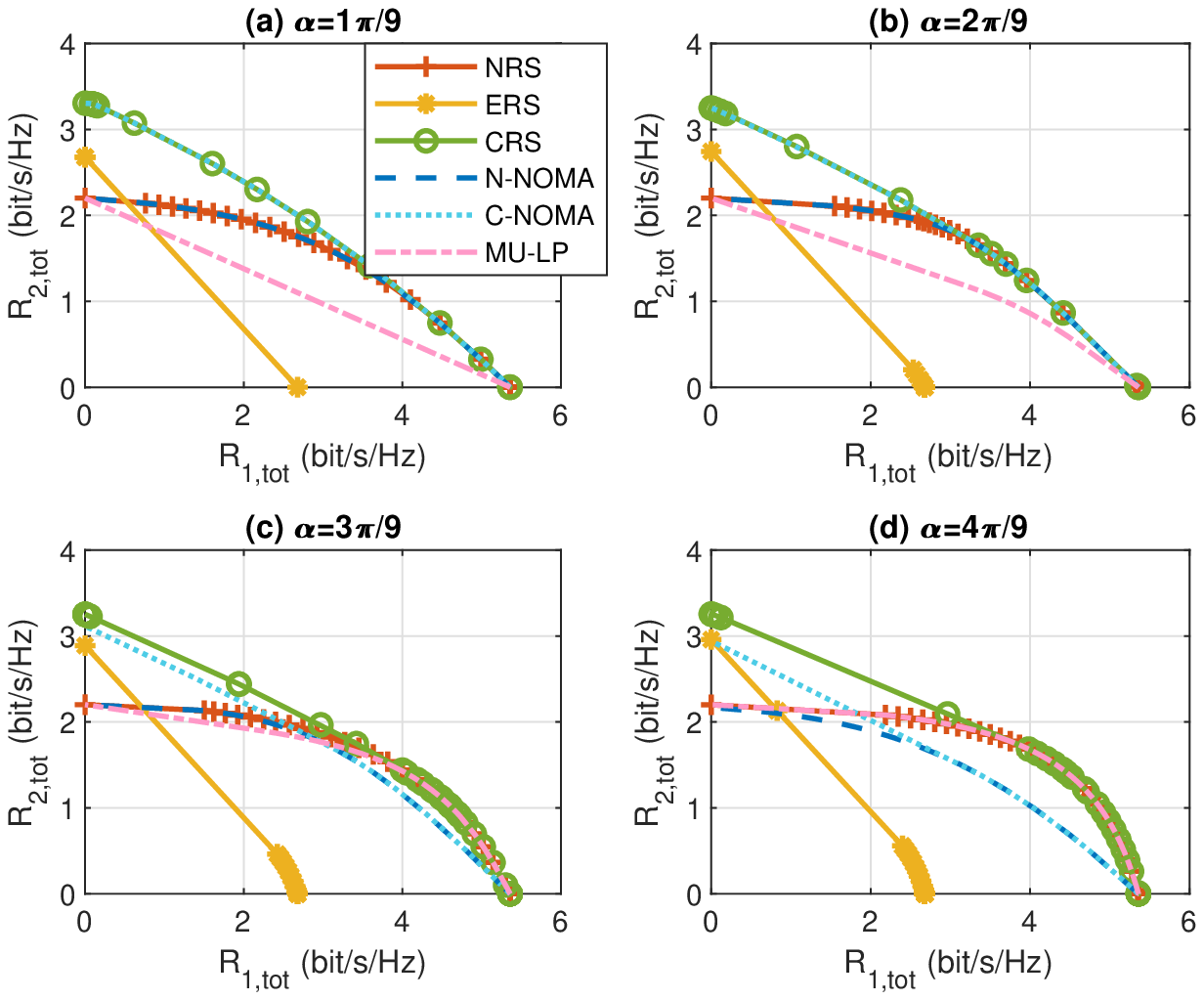}
	\vspace{-0.9cm}
	\caption{Achievable rate region comparison of different schemes with ${N_T} = 4$,  $\text{SNR} = 10\text{dB}$, and ${\left\| {{{\bf{h}}_1}} \right\|}:{\left\| {{{\bf{h}}_{\rm{2}}}} \right\|}:\left| {{h_3}} \right| = 1:0.3:1$.}
	\label{fig3}
	\vspace{-3mm}
\end{figure}

\begin{figure}[!h]
		\vspace{-1mm}
	\centering
	\vspace{-0.05cm}
	\includegraphics[width=1\linewidth]{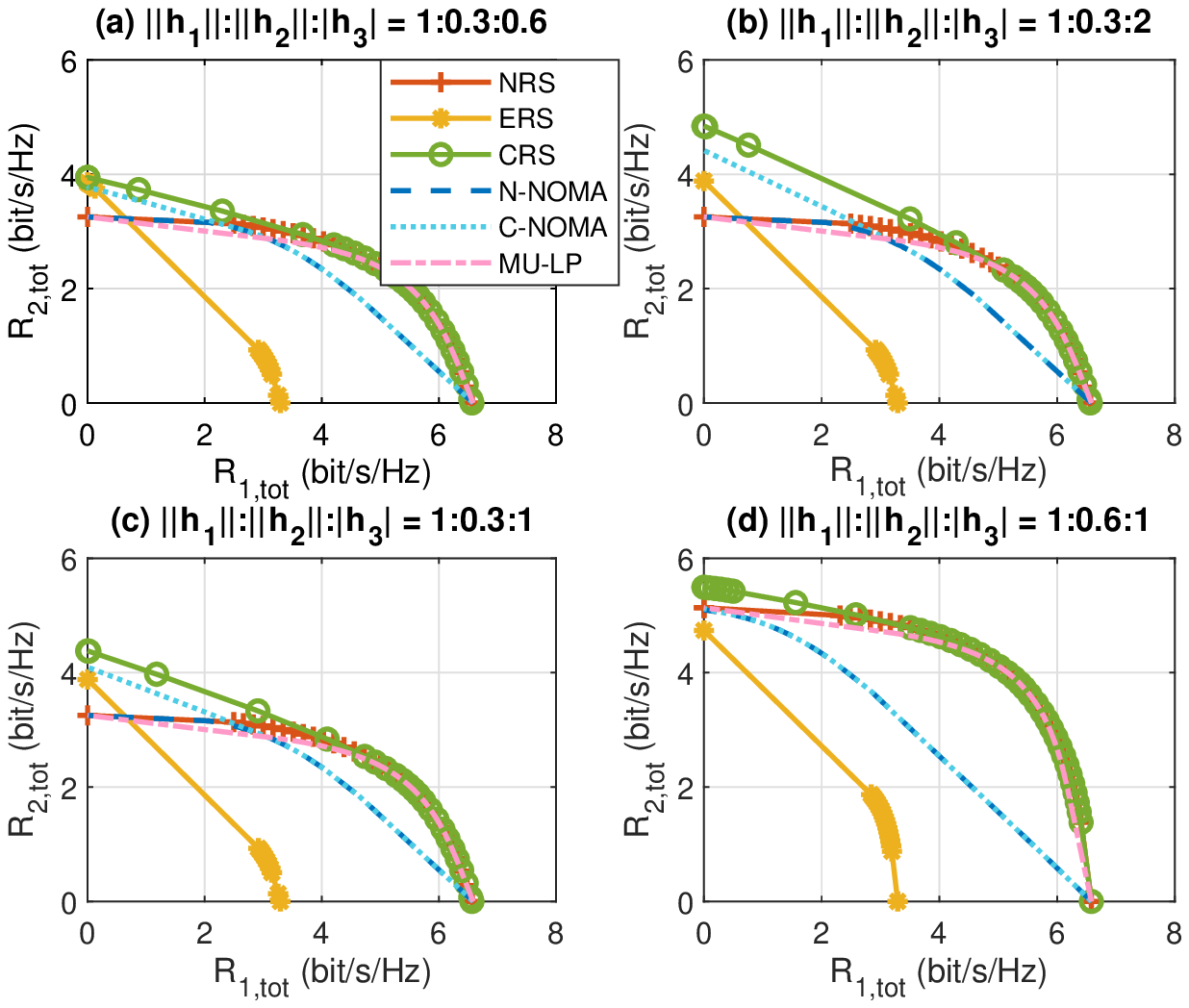}
	\vspace{-0.9cm}
	\caption{Achievable rate region comparison of different schemes with ${N_T} = 3$, $\text{SNR} = 15\text{dB}$, and ${\alpha} = 4\pi/9$.}
	\label{fig4}
	\vspace{-0.6cm}
\end{figure}

In Fig. 2, we first illustrate the rate region by varying the channel angle (from closely aligned to semi-orthogonal). In each subfigure, it is clear that the rate region achieved by CRS is equal to or larger than the other baseline schemes. 
Specifically, CRS reduces to NRS whenever $\theta$ is set to 1.
		As a higher priority is given to $U_2$, CRS brings benefits by performing user cooperation (or reducing $\theta$). In the extreme case when $u_2=10^3$ (at the edge point on $R_\text{2,tot}$ axis), CRS operates with a relative small $\theta$ and achieves the largest gain compared to NRS. Then, by comparing two cooperative strategies CRS and C-NOMA with their non-cooperative counterparts NRS and N-NOMA, we observe that the advantage of a relay link is  better exploited when the user channels are closely aligned. 
		This is due to the fact that the inter-user interference is much more severe when the user channels are aligned. By transmitting a relatively large amount of information via the common stream and using the relaying link to retransmit the common stream, the system performance is further enhanced in such case.
Furthermore, by comparing CRS with C-NOMA, we can observe that the rate region gap between the two strategies enlarges as $\alpha$ increases, a similar pattern is also obtained when comparing NRS and N-NOMA. This shows that the NOMA-based strategies are suited when users' channel directions are closely aligned, while RS-based strategies perform well in any channel angles.    

To further analyse the superiority of CRS and the interplay of different system parameters, we evaluate the rate region by varying the relative channel strength in Fig. 3. In all subfigures,
it is clear that CRS retains a certain gain over other schemes. 
As for ERS, due to the lack of time-domain flexibility, the rate region improvement of ERS over NRS is relatively limited and only appears in a few cases.
Comparing Fig. 3(a) with Fig. 3(b), it is not surprising that as the relay channel quality  $|h_3|$ improves, CRS and C-NOMA  become more beneficial over NRS and N-NOMA, which is a  general behaviour of relay system as in \cite{Liang2005,Liang2007,Gunduz2007,Liang2007a,Liang2007b}. Then, in Fig. 3(c) and Fig. 3(d), the rate region gap between CRS and C-NOMA enlarges as $\|{{{\bf{h}}_2}}\|$ increases from 0.3 to 0.6. The reason is that NOMA is motivated by leveraging the channel strength difference, and is sensitive to the channel strength disparity, while RS-based strategies are able to achieve good performance in all deployments.

\section{Conclusions}
In this letter, by naturally integrating the RS technique with classical three-node relay channel, we have proposed a more flexible and powerful cooperative scheme for MISO BC scenario. The precoder design and resource allocation problem were solved. Numerical results have demonstrated that CRS retains all the merits of RS, and outperforms other cooperative strategies like cooperative NOMA. Such promising properties therefore make the proposed scheme attractive for practical implementation in a wide range of scenarios.

\vspace{0.2cm}	
\bibliography{seventh-double}
\bibliographystyle{IEEEtr}

\end{document}